\documentclass[11pt,a4paper]{article}
\usepackage[utf8]{inputenc}
\usepackage[T1]{fontenc}
\usepackage{lmodern}
\usepackage{microtype}
\usepackage[margin=2.5cm]{geometry}
\usepackage{setspace}
\usepackage{parskip}
\usepackage{graphicx}
\usepackage{booktabs}
\usepackage{enumitem}
\usepackage{hyperref}
\hypersetup{colorlinks=true,linkcolor=blue,citecolor=blue,urlcolor=blue}
\usepackage{titling}
\usepackage{fancyhdr}
\usepackage{caption}
\usepackage{amsmath,amssymb}
\usepackage{wrapfig}
\usepackage{setspace}

\newcommand\gsim{\mathrel{\hbox{\rlap{\lower.55ex \hbox {$\sim$}}
                   \kern-.3em \raise.4ex \hbox{$>$}}}}
\newcommand\lsim{\mathrel{\hbox{\rlap{\lower.55ex \hbox {$\sim$}}
                   \kern-.3em \raise.4ex \hbox{$<$}}}}

\pretitle{\vspace*{-1cm}\begin{center}\LARGE\bfseries\end}\posttitle{\par\end{center}\vskip 0.5em}
\preauthor{\begin{center}\large}
\postauthor{\end{center}}
\predate{\begin{center}\large}
\postdate{\end{center}}

\begin{document}

\begin{titlepage}
  \centering
  \vspace*{1cm}
  {\Huge\bfseries Expanding Horizons \\[6pt] \Large Transforming Astronomy in the 2040s \par}
  \vspace{1.5cm}

  {\LARGE \textbf{Underluminous Thermonuclear Supernovae}\par}
  \vspace{1cm}

  \begin{tabular}{p{4.5cm}p{10cm}}
    \textbf{Scientific Categories:} & Stellar Evolution; Time-domain; Transients; Supernovae; Cosmology \\
    \\
    \textbf{Submitting Author:} & Name: Paul Groot \\
    & Affiliation: Radboud University (NL), University of Cape Town (ZA), South African Astronomical Observatory (ZA) \\
    & Email: p.groot@astro.ru.nl\\
    \\
    \textbf{Contributing authors:} & Simone Scaringi, Durham University, UK\\
    & Nancy Elias-Rosa, INAF, IT\\
  \end{tabular}

  \vspace{1cm}

  \textbf{Abstract:}

  \vspace{0.5em}
  \begin{minipage}{0.9\textwidth}
  Underluminous Thermonuclear Supernovae (uTSNe) are an emerging class of transient events that resemble classic Supernovae Type Ia, but peak at much lower luminosities. Suspected to be the deflagrations of white dwarfs, they directly link the final stages of low-mass binary star evolution to extragalactic studies that are critical for cosmology. The ability to detect and study uTSNe is limited by the lack of high spatial resolution ($<0.1^{\prime\prime}$), wide-field ($>10^\prime \times 10^\prime$) imaging capabilities in the optical, as well as large-scale segmented-telescope spectroscopic abilities that allow highly dynamic time-critical spectroscopy of short-duration transient events. Neither capability is currently foreseen for the European Southern Observatory and is therefore an excellent candidate for the Expanding Horizons program.
  \end{minipage}

\end{titlepage}


\section{Introduction and Background}
\label{sec:intro}


Thermonuclear supernovae (SN Ia) are used extensively to study the expansion history of the Universe and are a key ingredient in tracing cosmic acceleration [e.g. 1,2]. The lack of understanding on the exact origin and progenitor environment of thermonuclear supernovae is a continued worry underlying studies using SN Ia as tracers. Thermonuclear supernovae are the main source of intermediate-mass, iron-peak elements in the Universe and play an important role  in the chemical enrichment history of the Universe, providing elements that are essential for life.

These supernovae are generally accepted to correspond to the thermonuclear disruption of a white dwarf below or near the Chandrasekhar mass [e.g. 3]. They are powered by the fusion energy of (mainly) carbon to oxygen and further steps, which releases more energy than the white dwarf gravitational binding energy and causes a disruption. The trigger for nuclear fusion to start has to be mass gained, either through a merger event, or through accretion from a companion. This mass gain subsequently triggers fusion conditions, either in the core of the white dwarf or in an off-center location [e.g. 4-10].  

\begin{wrapfigure}{r}{0.49\textwidth}
\begin{center}
	\includegraphics[width=0.47\textwidth]{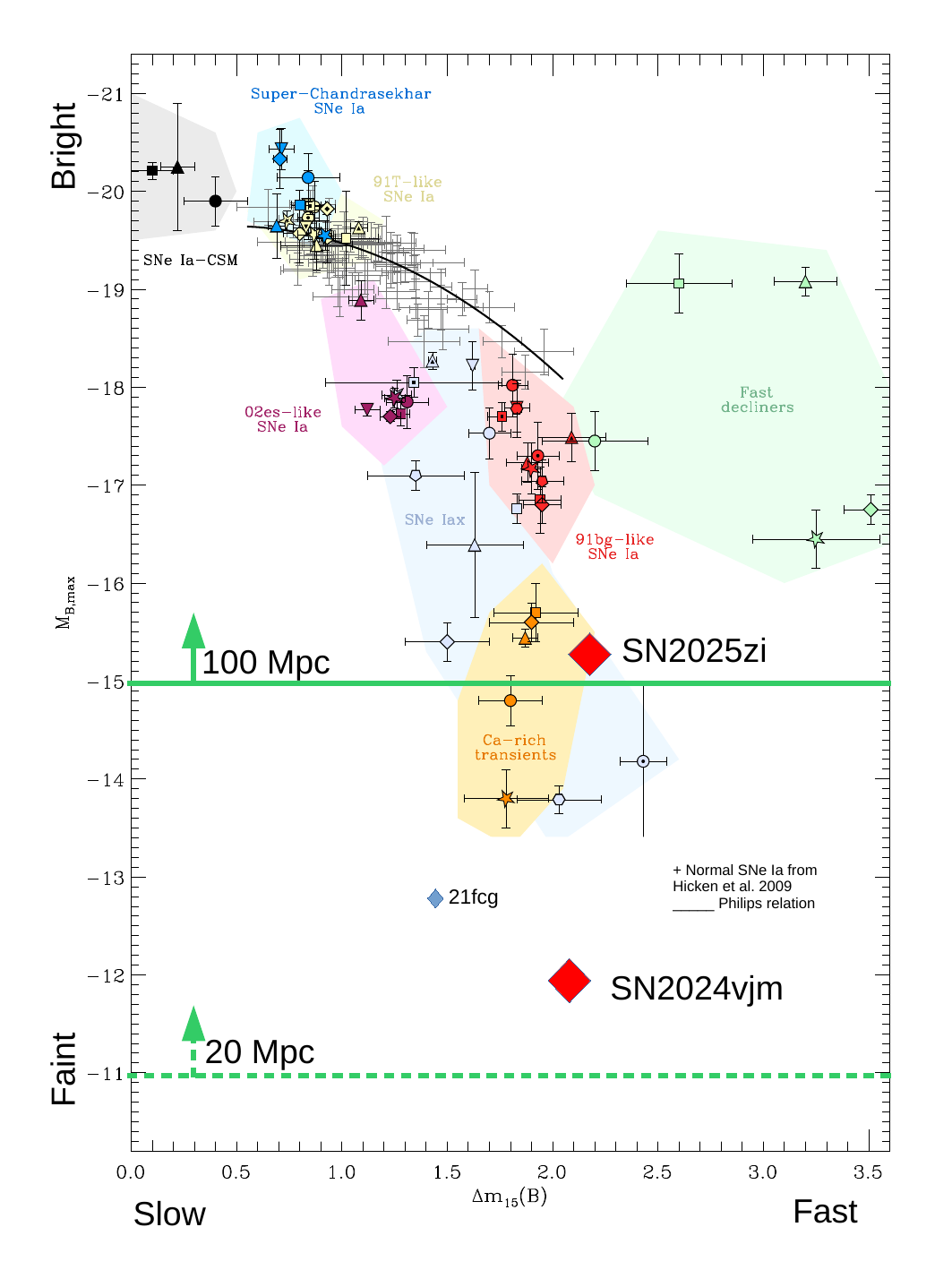}
    \caption{\scriptsize{Decay rate vs peak luminosity of thermonuclear supernovae, showing the classic SN Ia relation at the bright end and the growing population of underluminous supernovae ($M_{\rm B,max} < -18$). Luminosity limits as a function of distance are shown for 20 and 100 Mpc, assuming a detection limit equal to the sky background in the optical. Adapted from [11]}}
    \label{fig:uTSN}
    \end{center}
\end{wrapfigure}

Thermonuclear supernovae are highly luminous ($M_B \sim -19$), fully hydrogen-deficient explosions that are calibratable as cosmological standard candles. In the last decade a growing diversity of thermonuclear explosions that are clearly related to SN Ia have been detected [11-18]. These include SNe Iax, the `02es-like' Type Ia's and the Ca-rich transients (Fig.\ \ref{fig:uTSN}). In spectral appearance they are very similar to classic SN Ia, with the addition of stronger (IR) Ca lines, as well as lower outflow velocities, sometimes as low as only a few 1000 km s$^{-1}$. They are intrinsically lower luminosity at peak, ranging from $M_B = -12$ in the most extreme cases, up to the traditional limit of $M_B \lsim -18$ of SN Ia's. Observational characteristics point to incomplete white dwarf explosions as the origin of these {\sl underluminous thermonuclear supernovae (uTSN)}. Carbon fusion does not lead to a transition to a supersonic all-consuming detonation but remains as a deflagration, which burns part of the white dwarf, possibly leaving a bound remnant, hence leading to a lower energy, lower peak luminosity and lower outflow velocity, while maintaining a largely similar composition of the ejecta [5].

The physics, the progenitor setting and the evolutionary link to e.g. LISA sources, the rates of uTSN, and the relation with classic SN Ia are currently still unclear. Indications are that rates are possibly as high as 30\% of the SN Ia rate per galaxy [17], but this is still plagued by low number statistics as well as a strongly limited ability to detect the faintest events. A recent example of an extreme member of the class is SN2024vjm, which was found to have a peak luminosity of only $M_V = -12$, more than 500 times fainter than a classic SN Ia [19-20]. SN2024vjm was found because of its location in the very nearby face-on spiral galaxy NGC6744. Very deep pre-explosion {\sl Euclid} imaging allowed a high luminosity, giant companion/progenitor to be ruled out [20]. This shows how current observational limitations lead to a (possibly severe) underestimate on the rate and understanding of uTSNe. 

\vspace{-2mm}
\subsection{Developments in the coming years}
 uTSNe are primarily discovered from optical transient surveys, a rapidly growing field. In the coming five years, from the ground, the combined forces of ZTF [21], ATLAS [22], Pan-STARRS [23], GOTO [24], BlackGEM/MeerLICHT [25], and LAST [26], will be joined by LS4 [27], VRO/LSST [28] and on a longer time scale GOTTA [29], Argus [30] and the Digital Telescope[31]. Two general developments are: shorter time scales and deeper flux limits. Broadly speaking current surveys (ZTF, ATLAS, GOTO, BlackGEM/MeerLICHT, LAST) and the future LS4, Argus and the Digital Telescope surveys, are operating in natural seeing ($>$1$^{\prime\prime}$) and are limited by the sky background, which is g$\sim$21 mag/sq.arcsec in the mid-optical regime. Combined, these surveys already probe the full night sky at least once a night, providing a nightly cadence to the sky background limit. The Pan-STARRS survey probes deeper than the sky background and VRO/LSST is expected to go much deeper, up to 25th magnitude per visit, but at a low 3-night cadence per (changing) filter, or once per 15 days for a given filter. 

  Going to all-sky {\bf hourly cadences} is a relatively cheap option, since large format CMOS detectors and fast-slewing direct drive mounts are now affordable. This ties in with development routes taken by e.g. Argus (very large instantaneous field of view) and Digital Telescope (fixed mount, short integrations and dynamic stacking). In a decade the full night sky will be imaged down to the sky background limit on an hourly basis. For the uTSN this will allow both an earlier detection as well as a better sampling of the lightcurve for the few days they are bright enough. 

  Going {\bf deeper} will require additional developments. As uTSN are intrinsically faint events superimposed on high stellar backgrounds, a real rate determination not only requires going faster and to the sky background limit, but also a much higher spatial resolution to either resolve out the general sky background and make the fainter objects stand out earlier or further away; or to resolve out the stellar background glare in crowding-limited regions in the bulge or spiral arms of resolved galaxies. Even in VRO/LSST the actual limiting magnitude on nearby galaxies, including the Milky Way, will be set by stellar crowding limits and not the sky background. {\bf To beat the crowding limits wider-field optical adaptive optics will be needed to detect lower luminosity events. }

  In addition to high-resolution imaging, (time-resolved) spectroscopy will remain the main bottle neck. Routine spectroscopy of transient events is limited to the sky background regime, also in the era of the ELT. Currently only 10\% of all detected transients gets even a single spectrum [32]. An additional 10\% of the 10\%, i.e. only 1\% gets either a detailed multi-colour light curve or a time-resolved spectroscopic study [33]. Therefore, 90\% of the total transient sample remains spectroscopically unidentified and 99\% of all transients lack a detailed study. The NTT/SOXS combination [34] will offer some respite but, at a rate of 10-20 spectra per night, doesn't allow the aperture/resolution combination to solve the 99\% issue. {\bf A highly flexible, multi-aperture, multi-resolution spectroscopic array will be required}. A system that will allow an on-demand dynamic scheduling of tens to hundreds of targets, spread out over the night sky, with a configurable aperture setting to obtain repeated low/medium/high-resolution spectroscopy of transients at phases in their evolution where spectroscopic information is needed. We stress the time-critical nature of the required scheduling.

\vspace{-2mm}
\section{Open Science Questions in the 2040s}
\label{sec:openquestions}
We expect the field of uTSNe in the 2040s to be driven by the following questions: \\
    $\bullet$ What are the rates of uTSNe as a function of their host population (young/old stellar populations, high/low metallicities) and progenitor systems? \\
    $\bullet$ What is the intrinsic rate of uTSNe as a function of peak/bolometric luminosity and total energy and how do they connect up to binary white dwarf mergers?\\
    $\bullet$ What is the relation of uTSNe to classic SN Type Ia, and which class is caused by which explosion mechanism and progenitor setting? Is there a bound remnant surviving the explosion?\\
    $\bullet$ What is the contribution uTSNe to the chemical enrichment of the Universe? \\
    $\bullet$ Is there a systematic effect of the explosion mechanism(s) of (u)TSNe to their applicability as cosmological markers. 

\vspace{-2mm}
\section{Technology and Data Handling Requirements}
\label{sec:tech}
Required technological developments for the study of uTSN are: \\
    ${\bf 1)}$ Routine Adaptive Optics (AO) assisted wide-field ($> 10^\prime \times 10^\prime$) imaging in the optical (0.4 - 1 $\mu$m) with photometric stability on 1-4m class telescopes. This is required to obtain $\sim 0.1^{\prime\prime}$ resolution synoptic survey capabilities to resolve out high stellar backgrounds in the nearby Universe and be able to detect intrinsically low-luminosity transients. \\
    ${\bf 2)}$ A highly dynamic, reconfigurable spectroscopic array consisting of $>$100, AO-enabled 1-4m class telescopes, using a smaller number of fiber-fed optical-infrared spectrographs, spread across the globe. This will require zero-read-out-noise detectors, routine visible AO corrections on small fields in the optical, and a highly dynamic scheduler system that allows for on-demand `spectra-when-you-need-them' observations of targets with an intriniscally low sky surface density: $<$ 1 per square degree {\it at any given time} in the critical evolution of an event.

{\bf Neither capability is foreseen for ESO}: VLT/MAVIS will have a field-of-view of (only) 30$^{\prime\prime} \times 30^{\prime\prime}$ and ELT/MICADO even smaller. The VLT offers at most a 4-unit segmented telescope system, and the ELT is poorly suited for regular time-critical time-domain astronomy. 
The time critical aspect of the required observations, as well as the direct availability of data products, needs to be stressed. By 2040 the field of time-domain, transient, astronomy will have moved to a cadence shorter than 1 hour across the whole sky, and therefore follow-up facilities, reaction times and scheduling capabilities will also need to be on these time scales.



%

{\scriptsize
1: Riess et al., 2022, ApJL 934, L7;
2: Aghanim et al., 2018, A\&A 641, 6;
3: Ruiter \& Seitenzahl, 2025, A\&ARv 31, 1;
4: Bauer et al., 2019, ApJ 887, 68;
5: Fink et al., 2014, MNRAS 438, 1762;
6: Nomoto, 1982, ApJ 253, 798;
7: Pakmor, et al., 2022, MNRAS 517, 5260;
8: Shen, 2025, ApJ 982, 6;
9: Tanikawa, et al., 2019, ApJ 885, 103;
10: Wong et al., 2024, ApJ 973, 65;
11: Taubenberger, 2017, in Handbook of Supernovae, 317;
12: White et al., 2015, ApJ 799, 52;
13: Perets et al., 2010, Nature, 465, 322;
14: Kasliwal et al., 2012, ApJ 124, 854;
15: Leibundgut et al., 1993, AJ 105, 301;
16: Jha, 2017, in Handbook of Supernovae, 375;
17: Foley et al., 2013, ApJ 767, 57;
18: Filippenko et al., 1992, AJ 104, 1543;
19: Groot et al., 2024, TNSAN 255 1;
20: Zimmerman et al., 2025, in prep.;
21: Bellm, et al., 2019, PASP 131, 8002;
22: Tonry et a;., 2018, PASP 130, 4505;
23: Chambers et al., 2016, ArXiv, 61205560;
24: Steeghs et al., 2022, MNRAS 511, 2405;
35: Groot et al., 2024, PASP, 136, 5003;
26: Ofek et al., 2023, PASP, 135, 4502;
27: Miller et al., 2025, PASP 137, 4204;
28: Ivezic et al., 2019, ApJ 873, 111;
29: Liu et al., 2021, AnABC, 93, 628;
30: Law et al., 2022, PASP, 134, 5003;
31: Pollacco et al., : https://warwick.ac.uk/fac/sci/physics/research/astro/research/digit/;
32: Kulkarni, 2020, ArXiv 2004:03511;
33: Mogawana et al.,2025, A\&A, {\sl in prep};
34: Shipani et al., 2018, SPIE 10702;
}

\end{document}